\documentclass[10pt, conference, compsocconf]{IEEEtran}
\def\IEEEbibitemsep{0pt plus .5pt}
\usepackage{comment}
\usepackage{cite}
\usepackage{amsmath}
\usepackage{graphicx}

\ifCLASSINFOpdf
\else
\fi
\begin{document}
\title{Beyond Classification: Latent User Interests Profiling from Visual Contents Analysis}


\author{\IEEEauthorblockN{Longqi Yang}
\IEEEauthorblockA{Department of Computer Science\\
Cornell Tech\\
New York, USA\\
Email: ylongqi@cs.cornell.edu}
\and
\IEEEauthorblockN{Cheng-Kang Hsieh}
\IEEEauthorblockA{Department of Computer Science\\
University of California, Los Angeles\\
Los Angeles, USA\\
Email: changun@cs.ucla.edu}
\and
\IEEEauthorblockN{Deborah Estrin}
\IEEEauthorblockA{Department of Computer Science\\
Cornell Tech\\
New York, USA\\
Email: destrin@cs.cornell.edu}}


%Telephone: (800) 555--1212, Fax: (888) 555--1212}


\maketitle

\begin{abstract}
%\boldmath
%as it provides opportunities for personalization and effective targeting advertisements.
User preference profiling is an important task in modern online social networks (OSN). With the proliferation of image-centric social platforms, such as Pinterest, visual contents have become one of the most informative data streams for understanding user preferences. Traditional approaches usually treat visual content analysis as a general classification problem where one or more labels are assigned to each image. Although such an approach simplifies the process of image analysis, it misses the rich context and visual cues that play an important role in people's perception of images. In this paper, we explore the possibilities of learning a user's latent visual preferences directly from image contents. We propose a distance metric learning method based on Deep Convolutional Neural Networks (CNN) to directly extract similarity information from visual contents and use the derived distance metric to mine individual users' fine-grained visual preferences. Through our preliminary experiments using data from 5,790 Pinterest users, we show that even for the images within the same category, each user possesses distinct and individually-identifiable visual preferences that are consistent over their lifetime. Our results underscore the untapped potential of finer-grained visual preference profiling in understanding users' preferences.
\end{abstract}


%\IEEEpeerreviewmaketitle

\section{Introduction}
With the increasing popularity of different online social platforms, such as Facebook, Twitter, Pinterest etc., multi-modal data streams (e.g. text, image, audio, video, etc) are generated as byproducts of people's everyday online activities in the digital world. The wide availability of these \textit{digital breadcrumbs} \cite{Estrin:2014:SDN:2580723.2580944} have already cultivated major research efforts in the industry and academia to develop techniques to understand personal preferences. These techniques have led to the success of recommendation systems \cite{das2007google, middleton2004ontological}, such as Yelp, Foursquare etc., that help users find things they will enjoy, and enabled accurate targeting of advertisements.
%%Add citation to a top/general paper or two on recommendation systems

\begin{figure}
\centering
\includegraphics[width=0.95\linewidth]{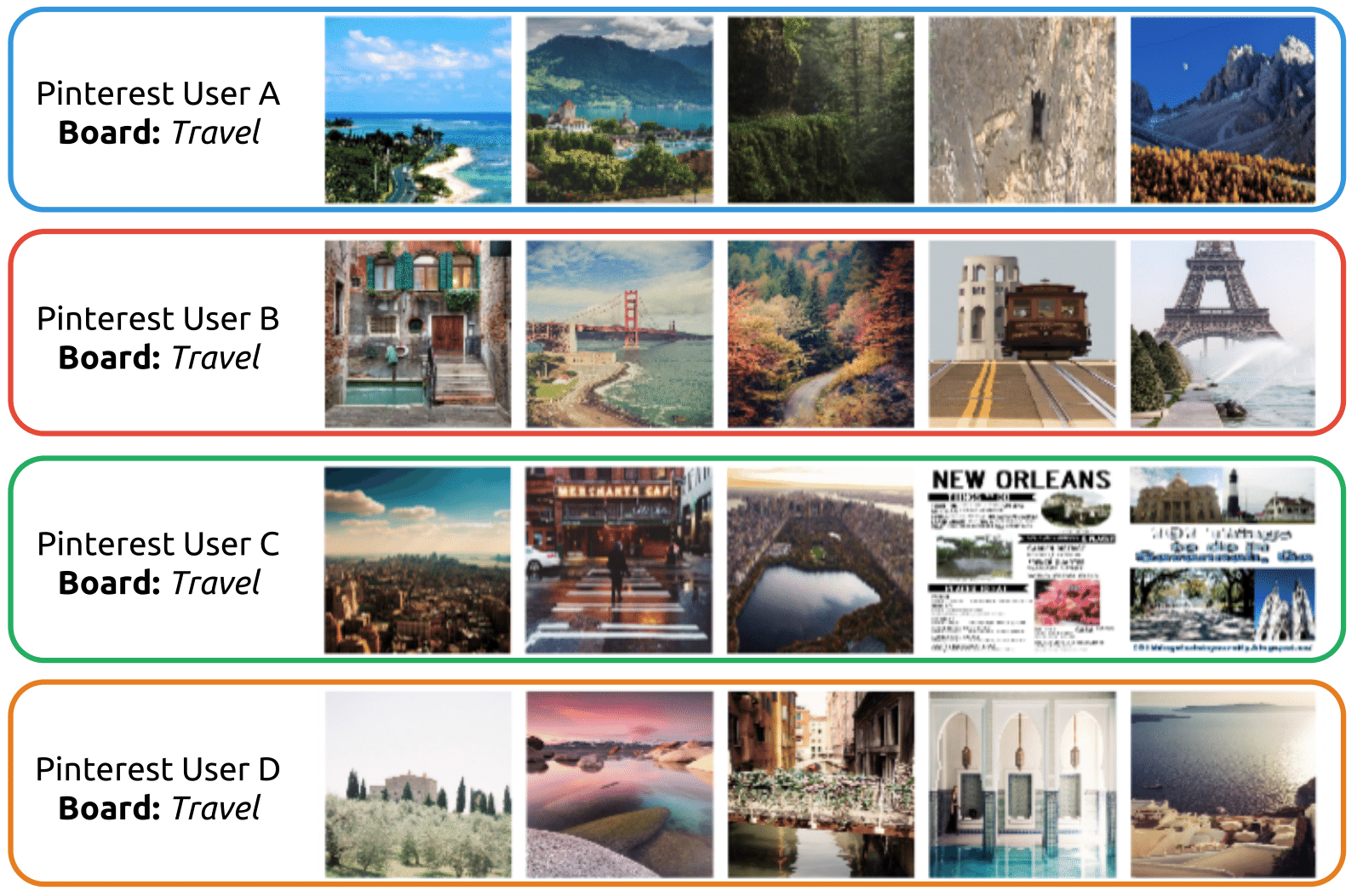}
\caption{Image samples from four travel boards curated by different users (All images are chronologically randomly sampled from users' boards)}
\label{fig:sample}
\vspace{-5mm}
\end{figure}

Text-centric data, such as tweets, and status updates, are among the most popular data streams for profiling personal attributes \cite{schwartz2013personality} due to their early adoption and pervasiveness. It has been shown by \cite{ schwartz2013personality, correa2010interacts, bamman2014gender} that various personal traits, such as gender, age, extroversion and openness, are manifested in these language features. Until recently, as driven by the emergence of photo sharing social media sites (e.g. Pinterest and Instagram) and the wide availability of embedded cameras on mobile devices, images have become a significant portion of contents that people posted online, and text data is thus limited for not capturing visual preferences. %% (Done) say why text wasnt enuf and what limitations caused people to explore other modalities.
Building on this line of research, some recent work started to explore the value of visual contents in uncovering people's interests \cite{you2015picture, ottoni2014pins, lovato2013we, lovato2014faved}. However, most current research in this domain \cite{you2015picture, ottoni2014pins, lovato2013we} converts images to one or more labels, and uses the text-based, categorical information to understand users' preferences. While such image-to-text approaches can benefit from the existing techniques developed for text-based data, they potentially miss the rich context and visual cues that are known to affect and guide people's perceptions of image contents \cite{gibson1950perception}. This limitation is especially highlighted on image intensive social networks, such as Pinterest. For example, as Fig.\ref{fig:sample} shows, even under the same category, \textit{Travel}, there are obvious distinctions between the \textit{pins} (i.e. the images on Pinterest) curated by different users. These distinctions could play an important role in not only image recommendations itself, but also in domains, such as travel destination recommendations. 

In this paper, we take a step \textit{deeper} into profiling users' visual preferences for images under the \textbf{same} label. We propose a novel framework based on Deep Convolutional Neural Networks (CNN) to directly learn a image distance metric from a large set of similar and dissimilar image pairs. We then leverage this similarity measure to profile each user's visual preferences. The experimental results, based on 5,790 Pinterest users' pins under the Travel category, indicate that the proposed approach is able to reveal each user's distinct visual preferences, and the derived user profile has strong predictive power to predict the images that the user will \textit{pin}. %% (Done) overkill to say enjoy??? say 'select' or 'choose' ???.

Compared with traditional solutions, our work offers three major contributions:

\begin{itemize}
\item Our approach enables fine-grained user interest profiling directly from visual contents. For images under the same label, we reveal intra-categorical variances that traditional classification methods were not able to capture.

\item We propose a novel distance-metric learning method based on the combination of traditional-CNN and Siamese Network\cite{chopra2005learning} models. This framework outperforms the state-of-the-art CNN model in terms of mean Average Precision (mAP).

\item Our experiment demonstrates \textit{beyond classification} utilities of visual contents in user interest profiling. We believe that our findings, while preliminary, shed light on the potential of incorporating fine-grained visual content analysis as an important technique for personalization.
\end{itemize}

\section{Related Work}

\subsection{Visual Content Analysis on OSNs}
The pioneering work in this domain studied online photos on Flickr \cite{lovato2013we, lovato2014faved, schifanella15image} and demonstrated the feasibility of extracting aesthetic and biometric features from user-generated image collections. It has been shown that people's preferences over these photographic features are identifiable and could be used for personalization \cite{yeh2010personalized}. Building on these prior efforts, recent literature has begun to explore the possibilities of profiling user's behavior \cite{ottoni2014pins, zhong2013sharing, bernardini2014pin} and interests \cite{you2015picture} from visual contents posted on social media. Although the work from \cite{you2015picture} has shown the initial findings of intra-categorical image variations among different users, most existing approaches treated image analysis as a classification problem where one or more labels are assigned and processed in a manner similar to text data. The major limitation behind such approaches is that a general classification model is trained and applied to all the users while ignoring individual users' distinct perception and preferences to an image category. Our preliminary experiments show that individual users do have distinct preferences even under the same category, and this personal preference is consistent over the user's lifetime. %% (Done)This last sentence is a bit obvious/tautological.

\subsection{Image Retrieval and Personalization}
The algorithms we propose in this paper are related to the \textit{similar image retrieval} problem in computer vision \cite{kim2012web, deng2011hierarchical, fu2015tagging}, where given a text query, semantically relevant images will be returned from a large database. It's similar to our work because the image similarity metric is an important component of the retrieval function and it has been shown that the algorithmic performance will achieve major improvements when incorporating user interests profile and temporal patterns of social events\cite{kim2012web}. Although most retrieval functions directly use visual features for similarity measurement\cite{kim2012web, deng2011hierarchical}, it is still unclear whether images themselves could provide utilities other than categorical labels and the extent of their usefulness in personal interest profiling. In this paper, we conduct experiments using publicly available data from 5,790 Pinterest users. The results demonstrate identifiable signals from visual contents that extend beyond classification and image categories. 
%%A rather random comment that you dont need to address for this paper is whether psychologists/neuroscientists have studied visual preference wrt art (eg paintings/drawings) where an individuals responses are often clearly to visual features/style more than subject matter...

\section{Problem Definition}
The general question we intend to answer in this paper is \textit{whether user-generated visual contents have predictive power for users' preferences beyond labels}. To quantitatively measure the differences of visual contents posted by different users under the same category, we consider the following setup of the problem.

Under an image category, each active user who posted in this category is denoted by $u_{i}, u_{i} \in \{u_{1}, u_{2}, ..., u_{N}\}$, and the images a user posted are denoted by $\mathcal{S}^{i}=\{I_{1}^{i}, I_{2}^{i}, ..., I_{\mid\mathcal{S}^{i} \mid}^{i}\}$ in the chronological order. The problem is to find a function $G$ such that $\boldsymbol{v}_{i} = G(\mathcal{S}^{i})$ can accurately characterize the user $i$'s distinct visual preferences. More specifically, we consider the following two tasks: 

(1) \textbf{Pairwise Comparison:} Given the general characteristics $\boldsymbol{\overline{v}}$ of images posted under this category, we analyze whether the proposed profiling function $G$ can distinguish the pairwise users' preferences so that the differences between each derived profile pair $(\boldsymbol{v}_{i}, \boldsymbol{v}_{j})$ are statistically significant.

(2) \textbf{Prediction:} We divide every user's image set $\mathcal{S}^{i}$ into training $(\mathcal{S}^{i}_{\text{train}})$ and testing $(\mathcal{S}^{i}_{\text{test} })$ subsets, and evaluate the predictive power of profile $\boldsymbol{v}_{i}^{\text{train}}$ by using it to predict which is the user $i$'s collections (board) among all the testing sets.

\section{Dataset Collection}
We choose Pinterest as the targeted platform since it is one of the most popular image-centric social networks. On Pinterest, users posted \textit{pins} (i.e. typically an image along with a short description) and organized them in self-defined \textit{boards}, each of which is associated with one of 34 predefined categories. This fully structured way of image collection makes Pinterest a natural candidate for investigating intra-categorical user preferences. In this paper, we scraped different users' boards within the \textit{travel} category. These travel boards are further filtered by the following two criteria: (1) The board should contain no less than 100 pins to guarantee that there is enough data for each user; and (2) The board should have at least one pin posted after June 2014 to ensure that the user is  still active\cite{danescu2013no}. After filtering, we obtained 5,790 \textit{travel boards}, each of which belongs to a different user. We use 1,800 of them as background corpus $\mathcal{S}_{\text{bg}}$ and exclude them from the analysis.

\section{Proposed Methodology}
\begin{figure*}[t]
\centering
\includegraphics[width=0.9\linewidth]{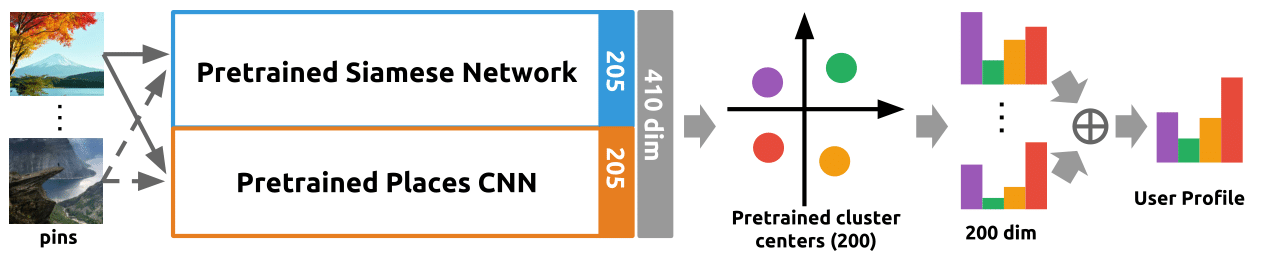}
\caption{Algorithmic framework for user interests profiling from visual contents. \textit{Phase 1:} Siamese Network and CNN based feature extraction; \textit{Phase 2:} Euclidean distance based soft assignment to pre-trained visual clusters; \textit{Phase 3:} Generate user profile by aggregating all image visual cluster features.}
\label{fig:overview}
\vspace{-2mm}
\end{figure*}

% that, given a user $u_{i}$, derive a profile for her travel visual interest $v_{i} = G(\mathcal{S}^{i})$ with all the pins in her travel board $\mathcal{S}^{i}=\{I_{1}^{i}, ..., I_{\mid \mathcal{S}^{i} \mid}^{i}\}$,

Fig.\ref{fig:overview} shows an overview of the proposed framework. The framework consists of three major components: (1) Each image (i.e. \textit{pin}) $I_{j}^{i}$ is first embedded in a 410-dimensional feature space via a pre-trained Siamese network and the Places-CNN. The feature vector for each image $I_{j}^{i}$ is denoted by $\boldsymbol{d}_{j}^{i}$; (2) Based on the distance between $\boldsymbol{d}_{j}^{i}$ and the center of each pre-trained visual cluster, an image is soft-assigned to 200 pre-trained clusters such that the final representation ($\boldsymbol{c}_{j}^{i}$) for the image $I_{j}^{i}$ is its affinities to all the clusters. (3) Finally, a user profile $\boldsymbol{v}_{i}$ is defined as the aggregate of all the feature vectors $\boldsymbol{c}_{1}^{i}, ..., \boldsymbol{c}_{\mid \mathcal{S}^{i} \mid}^{i}$. i.e. $\boldsymbol{v}_{i} = \frac{1}{Z} \sum_{j=1}^{\mid \mathcal{S}^{i} \mid} \boldsymbol{c}_{j}^{i}$, where $Z = \| \boldsymbol{v}_{i} \|_{1}$. In the following, we discuss important design decisions and the rationales behind each component.

\subsection{Deep Distance Metric Learning}
Distance metric learning using Deep Siamese Network has achieved significant performance improvements in face verification\cite{taigman2014deepface}, geo-localization\cite{lin2015learning} and food image embedding\cite{yang2015CIKM}. In addition, it is suggested by \cite{sun2013hybrid} that feature concatenation (hybrid) from CNNs trained under different conditions will further strengthen the discriminative power of the model. In light of these prior efforts, we fine-tuned a Siamese Network based on the Places dataset\cite{zhou2014learning} and concatenated its features with the pre-trained Places-CNN model\cite{zhou2014learning} (Fig.\ref{fig:overview}), both of which utilized the AlexNet\cite{krizhevsky2012imagenet} architecture. We choose to use the Places dataset and include the Places-CNN model because the images we deal with are mostly scene photos from the \textit{travel} category. In this section, we focus on our design and training choices for the Siamese Network. Interested readers can refer to the original papers for details\cite{krizhevsky2012imagenet}. 

\begin{figure}
\centering
\includegraphics[width=0.95\linewidth]{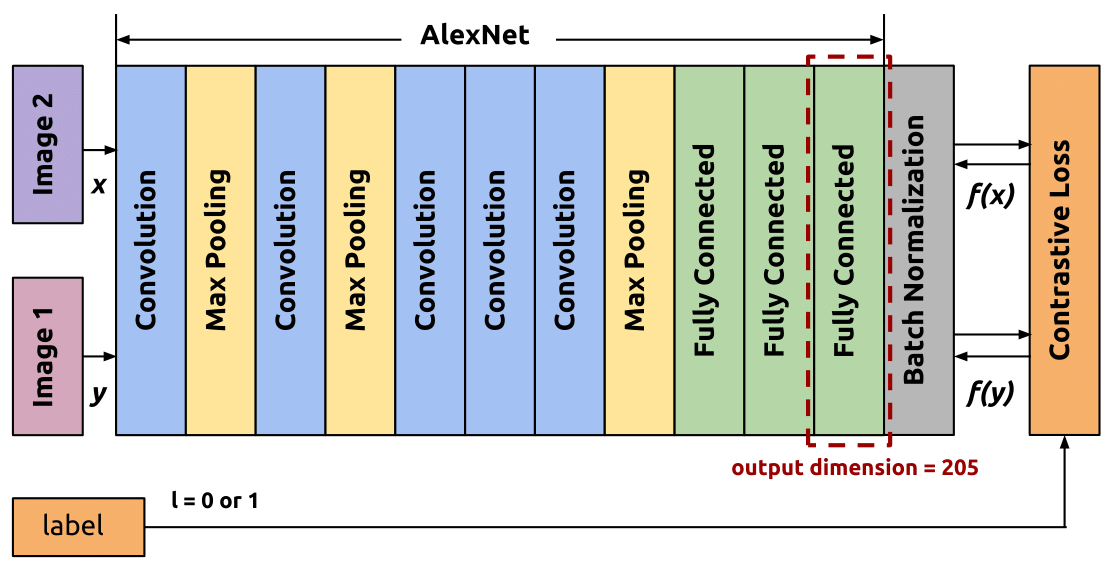}
\caption{Structure of Siamese Network used in the feature embedding}
\label{fig:siamese}
\vspace{-2mm}
\end{figure}

As illustrated in Fig. \ref{fig:siamese}, our Siamese Network architecture is the same as AlexNet\cite{krizhevsky2012imagenet} except that we change the output dimension of the last fully connected layer to 205 in order to stay consistent with the output of Places-CNN. We also add a Batch Normalization layer \cite{ioffe2015batch} at the end to normalize the 205 dimensional feature so that each dimension has zero mean and unit variance within a training batch.  Our goal is to learn a low dimensional feature embedding where similar scene images are pulled together while dissimilar images are pushed far away. Specifically,  we want $f(x)$ and $f(y)$ to have small distance (close to 0) if $x$ and $y$ are similar instances; otherwise, they should have distance larger than a margin $m$. In this paper, we choose Contrastive Loss $\mathcal{L}$ proposed in \cite{hadsell2006dimensionality} as the loss function when optimizing the Siamese Network.

\begin{equation}
\label{eqn_loss}
\mathcal{L}(x, y, l) = \frac{1}{2} l D^2 + \frac{1}{2} (1 - l) \max\left(0, m - D\right)^2
\end{equation}

In eqn.(\ref{eqn_loss}), similarity label $l \in \{0, 1\}$ indicates whether the input pair of scene images $x, y$ are similar or not ($l=1$ for similar, $l=0$ for dissimilar), $m>0$ is the margin for dissimilar scenes and $D=\|f(x) - f(y)\|_{2}$ is the Euclidean Distance between $f(x)$ and $f(y)$ in the embedding space. We use the open-source implementation of gradient descent and back-propagation provided by Caffe\cite{jia2014caffe} to train and test Siamese Network. 

In the training phase, we treat the Places dataset images with the same labels as similar pairs and those under different categories as dissimilar pairs. We sample 102,500 similar pairs and 1,045,500 dissimilar pairs to train our Siamese Network. We set the learning rate of the last fully connected layer as $10^{-5}$ and the rate for the rest layers as $10^{-7}$. The model that we use in this paper is trained for 50,000 iterations. Finally, the output of Siamese Network (205 dimension) will be concatenated with the output of the fully connected layer in Places-CNN, which together form a 410 dimensional feature embedding for each image.

\subsection{Clustering and User Profiling}
After the training phase, we use the pretrained Siamese Network and Places-CNN to extract 410 dimensional feature $\boldsymbol{d}_{j}^{i}$ for each image $I_{j}^{i}$. We randomly sample 1800 users and use their images $\mathcal{S}_{\text{bg}} = \mathcal{S}^{1}\cup...\cup\mathcal{S}^{1800}$ as the background corpus to discover latent clusters \footnote{They are excluded from the following pair-wise comparison and prediction tasks}. A traditional K-means\cite{macqueen1967some} unsupervised clustering algorithm is used to divide the image set into 200 visual clusters, and their centers are denoted by $\boldsymbol{r}_{1}, \boldsymbol{r}_{2}, ..., \boldsymbol{r}_{200}$. Built on the pre-trained cluster centers, each image is then soft assigned to 200 clusters based on eqn.(\ref{eqn:assignment}) such that each dimension of the final representation $\boldsymbol{c}_{j}^{i}$ reveals the likelihood of the image belonging to a specific visual cluster.

\begin{equation}
\boldsymbol{c}_{j}^{i}(k) = \left\{
     \begin{array}{lr}
       e^{-\frac{1}{2\alpha^{2}} \lVert \boldsymbol{d}_{j}^{i} - \boldsymbol{r}_{k} \rVert^{2}} & : \lVert \boldsymbol{d}_{j}^{i} - \boldsymbol{r}_{k} \rVert \leq \delta\\
       0 & : \lVert \boldsymbol{d}_{j}^{i} - \boldsymbol{r}_{k} \rVert > \delta
     \end{array}
   \right.
\label{eqn:assignment}
\end{equation}

where $\alpha^{2} = \frac{1}{\mid \mathcal{S}_{\text{bg}} \mid^{2}} \sum_{I_{j}^{i}, I_{n}^{l} \in \mathcal{S}_{\text{bg}}} \|\boldsymbol{d}_{j}^{i} - \boldsymbol{d}_{n}^{l} \|^{2}$ and $\delta = m$ ($m$ is the margin of Siamese Network).

Finally, for each user $u_{i}$, we derive her profile by aggregating all the image feature representations $\boldsymbol{c}_{j}^{i}$ in her collection of pins $\mathcal{S}^{i}$ via eqn.(\ref{eqn:aggregation}). This profile intuitively represents the distribution of users' interests over different visual clusters.

\begin{equation}
\boldsymbol{\tilde{v}}_{i} = \sum_{j=1}^{\mid \mathcal{S}^{i} \mid} \boldsymbol{c}_{j}^{i}; \quad \boldsymbol{v}_{i} = \frac{1}{\| \boldsymbol{\tilde{v}}_{i} \|_{1}} \boldsymbol{\tilde{v}}_{i}
\label{eqn:aggregation}
\end{equation}

\subsection{User Pairwise Comparison}
Given a pair of user $i$ and user $j$, we investigate whether the derived profile has the discriminative power to different users' preferences. Users' pairwise differences are evaluated over the general distribution $\boldsymbol{\overline{v}}$ of images under \textit{travel} boards. This general distribution is derived from the background corpus $\mathcal{S}^{\text{bg}}$, where $\boldsymbol{\overline{v}} = \sum_{I_{j}^{i} \in \mathcal{S}^{\text{bg}}} \boldsymbol{c}_{j}^{i}$. We adopt \textit{log odds ratio with informative Dirichlet prior} proposed in \cite{monroe2008fightin} to analyze pairwise differences; this approach was originally used for comparing the differences of word frequencies between articles.

We first calculate the \textit{log odds ratio} with respect to different visual cluster $k$ as in eqn.(\ref{eqn:log}), where $\alpha$ controls the size of background corpus.
%%ANDY: What do you mean by "controls"? is $alpha$=1800?

\begin{equation}
\begin{split}
\hat{\delta}_{k}^{\boldsymbol{v}_{i} - \boldsymbol{v}_{j}} &= log(\frac{\boldsymbol{\tilde{v}}_{i}(k) + \alpha\boldsymbol{\overline{v}}(k)}{\sum_{k}\boldsymbol{\tilde{v}}_{i}(k) + \alpha\sum_{k}\boldsymbol{\overline{v}}(k) - (\boldsymbol{v}_{i}(k) + \alpha\boldsymbol{\overline{v}}(k))}) \\
&- log(\frac{\boldsymbol{\tilde{v}}_{j}(k) + \alpha\boldsymbol{\overline{v}}(k)}{\sum_{k}\boldsymbol{\tilde{v}}_{j}(k) + \alpha\sum_{k}\boldsymbol{\overline{v}}(k) - (\boldsymbol{v}_{j}(k) + \alpha\boldsymbol{\overline{v}}(k))})
\end{split}
\label{eqn:log}
\end{equation}

In addition, we consider the estimated uncertainty as suggested in \cite{monroe2008fightin} and calculate the variance value as in eqn.(\ref{eqn:variance}). 

\begin{equation}
\sigma^{2}(\hat{\delta}_{k}^{\boldsymbol{v}_{i} - \boldsymbol{v}_{j}}) \approx \frac{1}{\boldsymbol{\tilde{v}}_{i}(k) + \alpha\boldsymbol{\overline{v}}(k)} + \frac{1}{\boldsymbol{\tilde{v}}_{j}(k) + \alpha\boldsymbol{\overline{v}}(k)}
\label{eqn:variance}
\end{equation}

The final statistic for each visual cluster $k$ is the z-score of the log-odds-ratio, computed as in eqn(\ref{eqn:zscore}).

\begin{equation}
z_{k} = \frac{\hat{\delta}_{k}^{\boldsymbol{v}_{i} - \boldsymbol{v}_{j}}}{\sqrt{\sigma^{2}(\hat{\delta}_{k}^{\boldsymbol{v}_{i} - \boldsymbol{v}_{j}})}}
\label{eqn:zscore}
\end{equation}

The method we adopt in this section takes into account the background corpus as prior, which alleviates the data sparsity problem and makes the differences of very frequent visual clusters detectable. Under such conditions, if $\mid z_{k}\mid \geq 2$, the confidence level that user $u_{i}$ and $u_{j}$ are significantly different is greater than $95\%$. We will show the overall distribution of all pairwise user differences in the following experiments section.

\section{Experiments}
\subsection{Distance Metric Evaluation}

\begin{figure}
\centering
\includegraphics[width=0.95\linewidth]{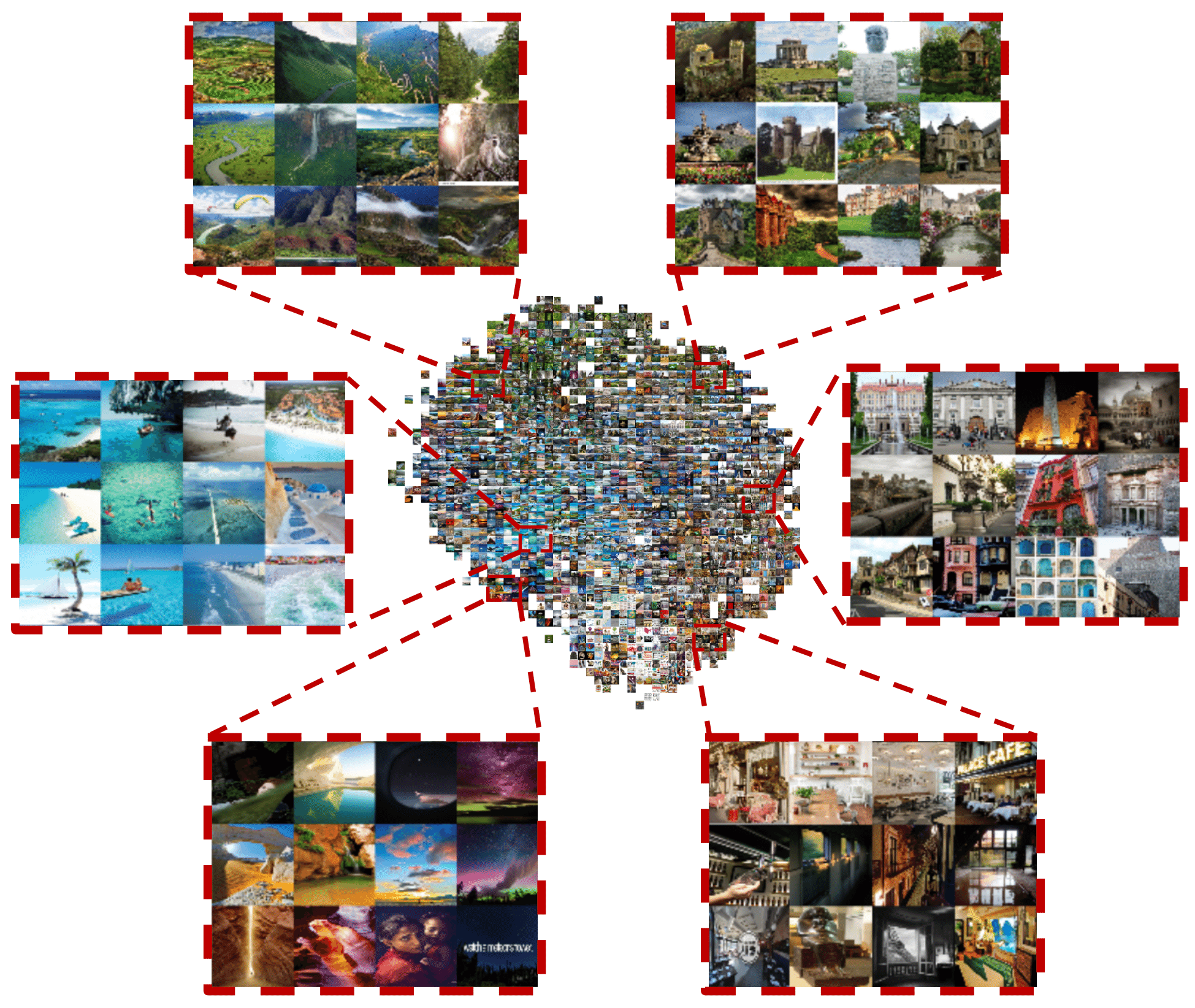}
\caption{Pinterest \textit{travel} images embedding based on our hybrid CNN model; The images are projected to 2-D plane using t-SNE.}
\label{fig:embedding}
\vspace{-3mm}
\end{figure}

\begin{table}[h]
\centering
\begin{tabular}{c|c|c|c}
\hline
\textbf{Hybrid CNN} & \textbf{Places CNN} & \textbf{SIFT+BOW} & \textbf{Random Guess} \\ \hline
\textbf{0.134}                            & 0.132           & 0.019
&  0.005                          \\ \hline
\end{tabular}
\vspace{3mm}
\caption{mean Average Precision (mAP) value of the image clustering task on Places dataset}
\label{tbl:map}
\vspace{-3mm}
\end{table}

We evaluate the efficacy of the distance metric derived from our hybrid model by measuring its clustering performance, namely to what extent the distance metric can cluster test images that share the same labels in the Places Dataset\cite{zhou2014learning}. We check the nearest $k$-neighbors of each test image for $k = 1, 2, ..., N$, where $N=20,500$ is the size of the testing dataset, and calculate the Precision and Recall values for each $k$. We use mean Average Precision (mAP) as the evaluation metric to compare the performance with the competing algorithms as suggested in \cite{yang2015CIKM}. For every method, the Precision/Recall values are averaged over all the images in the testing set. The results are shown in Table.\ref{tbl:map} where an ideal algorithm has mAP value equals to 1.%%(Done)reword this last sentence

We compare our hybrid model with two important competing algorithms: (1) \textit{Pretrained Places CNN}\cite{zhou2014learning}: We extract a 205-dimensional feature from the output of the last fully connected layer in the Places CNN and use it as the representation for each image; (2) \textit{SIFT+Bag of Words(BoW)}\cite{lowe2004distinctive}: For this state-of-the-art hand crafted representation, we extract features using 410 visual words so that it has the same feature dimension as our hybrid model. As is shown in Table.\ref{tbl:map}, traditional feature representation (\textit{SIFT + BOW}) does not have enough discriminative power for the task of scene image embedding. The hybrid model that we propose in this paper outperforms both of the approaches mentioned above in terms of mAP values. These evaluation results not only justify the value of the Siamese network method, but also show that the strategy of concatenating different CNN features could improve the performance of the model.

The feature embedding model proposed in this paper has the promise for visualizing and discovering image clusters among travel images. We randomly sample 10,000 pins from background corpus $\mathcal{S}_{\text{bg}}$ and project all images to a 2-D plane using t-Distributed Stochastic Neighbor Embedding (t-SNE)\cite{van2008visualizing}. As shown in Fig.\ref{fig:embedding}, we divide the plane into many small blocks, and for each block we randomly sample a representative scene image that resides in that area. The final embedding clearly groups similar scenes more closely in the new space. The embedding results (Fig.\ref{fig:embedding}) indicate that we can capture rather fine-grained image categories that are likely to appear in \textit{travel} boards. For instance, natural scenes (e.g. beach, mountains), city views (e.g. building, street) and travel necessities (e.g. bags, shoes).

\subsection{Pairwise Comparison}

\begin{figure}
\centering
\includegraphics[width=0.88\linewidth]{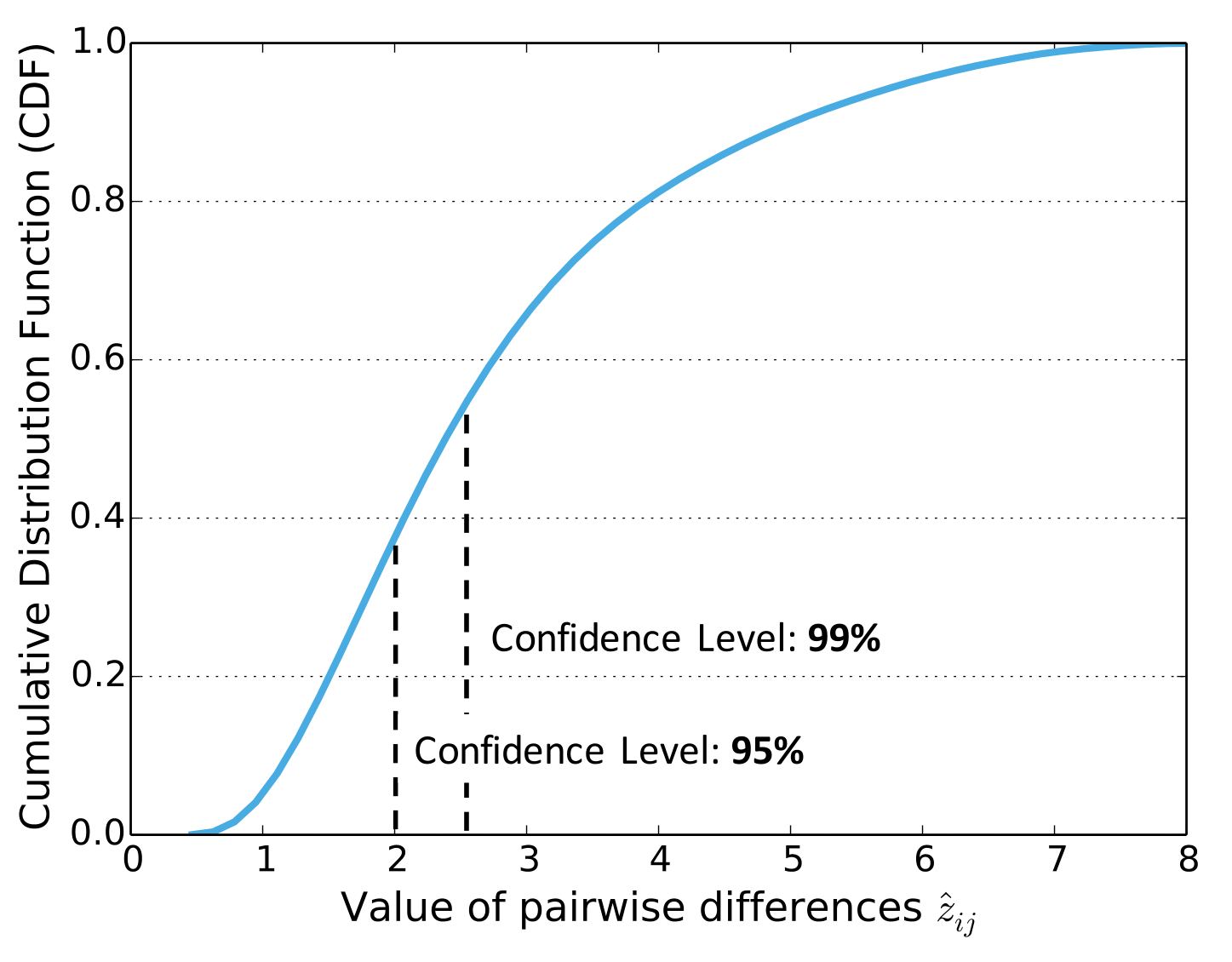} %that characterizes the differences between user pairs
\caption{Empirical Cumulative Distribution Function (eCDF) of $\hat{z}_{ij}$. The dotted lines denote the confidence levels associated with different $z$ scores. It shows that more than half of the user pairs have statistically significant differences (i.e. $\hat{z}_{ij} \geq 2$) in visual preferences even under the same category of images.}
\label{fig:cdf}
\vspace{-2mm}
\end{figure}

To investigate how much intra-categorical variance exists between Pinterest users, for each pair of users $(u_{i}, u_{j})$ (except those 1,800 users used for background corpus), we estimate the pairwise dissimilarity between them using the z-score described in Section V. More specifically, let $z_{ij,k}$ denote the z-score that estimates the difference between users $(u_{i}, u_{j})$ in the visual cluster $k$. Then, the overall preference difference between users $(u_{i}, u_{j})$, denoted by $\hat{z}_{ij}$, is estimated by the maximum z-score over all $K$ visual clusters as defined in eqn.(\ref{eqn:pair}).

\begin{equation}
\hat{z}_{ij} = \text{max}_{k} \mid z_{ij,k} \mid
\label{eqn:pair}
\end{equation}

We plot the empirical cumulative distribution function (eCDF) of $\hat{z}_{ij}$ for all the pairwise users in Fig.\ref{fig:cdf}. The distribution demonstrates that there are more than half of the user pairs that have statistically significant difference (i.e. $\hat{z}_{ij} \geq 2$) in their visual preferences even for the same category of images. This result verifies our assumption that there is significant intra-categorical variance among different users and underscores the importance of understanding users' fine-grained interests and preferences.

\subsection{Prediction of Future Pins Collections}

\begin{figure}
\centering
\includegraphics[width=0.9\linewidth]{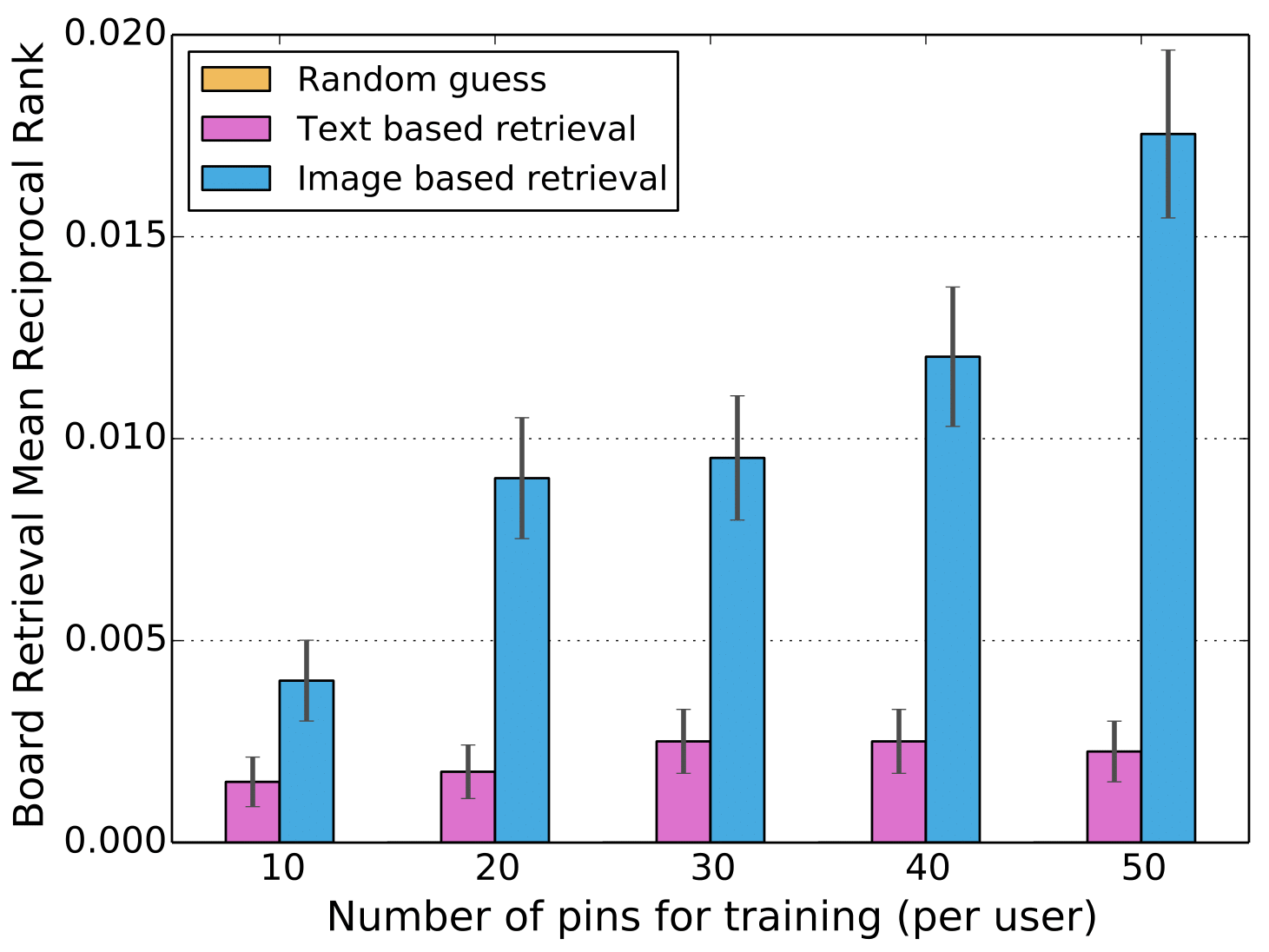}
\caption{Mean Reciprocal Rank (MRR) for the pin collection (i.e. board) retrieval task with different sizes of training samples. The performance is compared across three algorithms : \textit{Random guess}, \textit{Text similarity based retrieval} and \textit{Image similarity based retrieval}.}
\label{fig:retrieval}
\vspace{-6mm}
\end{figure}

In addition to pair-wise comparisons, the other question we want to answer is whether the user profile derived with our hybrid model has discriminative power to different users' preferences. In order to quantitatively measure that, we propose the following prediction task: (1) 100 images (denoted as $\tilde{\mathcal{S}}^{i}$) are randomly sampled from each image set $\mathcal{S}^{i}$ to guarantee that each user has the same number of pins for training and prediction; (2) Each sampled image set $\tilde{\mathcal{S}}^{i}$ is then divided into training ($\tilde{\mathcal{S}}^{i}_{\text{train}}$) and testing ($\tilde{\mathcal{S}}^{i}_{\text{test}}$) subsets based on their chronological order; (3) Each user's profile is calculated based on two sets separately (i.e. $\boldsymbol{v}_{i}^{\text{train}} = G(\tilde{\mathcal{S}}^{i}_{\text{train}}); \boldsymbol{v}_{i}^{\text{test}} = G(\tilde{\mathcal{S}}^{i}_{\text{test}})$); (4) For each user $i$ and her profile $\boldsymbol{v}_{i}^{\text{train}}$ based on her training set, we predict which testing set belongs to her using euclidean distances. More specifically, we sort all the testing sets $\tilde{\mathcal{S}}^{j}_{\text{test}}$ by the euclidean distances between their profile $\boldsymbol{v}_{j}^{\text{test}}$ and the user's profile $\boldsymbol{v}_{i}^{\text{train}}$ in an ascending order, and the ranking of the user's real testing set $\boldsymbol{v}_{i}^{\text{test}}$ is denoted as $rank_{i}$. Finally, Mean Reciprocal Rank (MRR), as defined in eqn.\ref{eqn:mrr}, is used to evaluate the overall prediction accuracy across all the users ($N=3,990$). MRR is a standard metric for evaluating the accuracy of a prediction algorithm.

\begin{equation}
\text{MRR} = \frac{1}{N} \sum_{i=1}^{N} \frac{1}{rank_{i}}
\label{eqn:mrr}
\end{equation}

In order to show the effects of the size of training set, we fix the testing set $\tilde{\mathcal{S}}^{i}_{\text{test}}$ to contain the last 50 pins in $\tilde{\mathcal{S}}^{i}$ and vary the training set $\tilde{\mathcal{S}}^{i}_{\text{train}}$ to include the first 10, 20, 30, 40, 50 pins. In addition, we compare our approach to a text-based user interesting profiling approach. The procedure for this text-based user interests profiling is similar to the one shown in Fig.\ref{fig:overview}, but, instead of using hybrid deep neural network, we adopt the state-of-the-art PV-DM model\cite{le2014distributed} to embed each pin's text description into a 100-dimensional feature space.

As is shown in Fig.\ref{fig:retrieval}, the profiles that we calculated based on visual contents have significantly better performance than text and random baselines in terms of Mean Reciprocal Rank. The results further demonstrate the possibilities that, in image-centric social networks (e.g. Pinterest), visual contents play a more significant role in affecting users' behavior and preferences compared to traditional text-based platforms. Although there is still a large space of algorithmic improvements to be explored, our preliminary results provide promising evidence for using intra-categorical variance information to understand people's interests and preferences.

\section{Future Work}
Moving forward, there are several directions we would like to pursue. (1) \textit{Comprehensive intra-categorical image analysis model:} in this paper, we only consider the images under the \textit{travel} category. However, in real world applications, there are a large number of image categories. A general and comprehensive model to analyze users' intra-categorical preferences for a wide variety of images categories will be of significant importance; (2) \textit{Information fusion of inter- and intra- categorical image analysis:} one of the opportunities enabled by the fine-grained image analysis is to fuse and propagate inter- and intra- categorical information. A hierarchical model could be built to analyze users' visual preferences in different levels and their inter-level interactions. Finally (3) \textit{cross-platform information sharing:} cross-platform behavior analysis is a user-centric idea to explore the sharing and fine-tuning of user profiles across multiple platforms. This will be particularly useful for solving cold-start problems \cite{park2006naive} in many recommender systems. For example, one can use users' fine-grained interests learned from Pinterest to recommend friends or places in another social network.

\section{Conclusion}
To conclude, in this paper, we propose a user preference profiling framework that extracts signals with strong discriminative power to users' fine-grained preferences. Compared to previous work, the proposed framework is a hybrid one that takes advantages of Siamese Network and traditional CNN to directly extract similarity information from images. Our experimental results based on data from 5,790 Pinterest users show that the proposed method is able to characterize the intra-categorical interests of a user with a resolution that is beyond what a coarse-grained image classification can do. Our findings suggest that there is great potential in finer-grained user visual preference profiling, and we hope this paper will fuel future development of deeper and finer understanding of users' latent preferences and interests.

\section*{Acknowledgement}
We appreciate the anonymous reviewers for insightful comments. This research is partly funded by AOL-Program for Connected Experiences and further supported by the small data lab at Cornell Tech which receives funding from UnitedHealth Group, Google, Pfizer, RWJF, NIH and NSF.


% \section*{Acknowledgment}



\bibliographystyle{IEEEtran}
\bibliography{IEEEexample}

\begin{thebibliography}{10}
\providecommand{\url}[1]{#1}
\csname url@samestyle\endcsname
\providecommand{\newblock}{\relax}
\providecommand{\bibinfo}[2]{#2}
\providecommand{\BIBentrySTDinterwordspacing}{\spaceskip=0pt\relax}
\providecommand{\BIBentryALTinterwordstretchfactor}{4}
\providecommand{\BIBentryALTinterwordspacing}{\spaceskip=\fontdimen2\font plus
\BIBentryALTinterwordstretchfactor\fontdimen3\font minus
  \fontdimen4\font\relax}
\providecommand{\BIBforeignlanguage}[2]{{%
\expandafter\ifx\csname l@#1\endcsname\relax
\typeout{** WARNING: IEEEtran.bst: No hyphenation pattern has been}%
\typeout{** loaded for the language `#1'. Using the pattern for}%
\typeout{** the default language instead.}%
\else
\language=\csname l@#1\endcsname
\fi
#2}}
\providecommand{\BIBdecl}{\relax}
\BIBdecl

\bibitem{Estrin:2014:SDN:2580723.2580944}
\BIBentryALTinterwordspacing
D.~Estrin, ``Small data, where n = me,'' \emph{Commun. ACM}, vol.~57, no.~4,
  pp. 32--34, Apr. 2014. [Online]. Available:
  \url{http://doi.acm.org/10.1145/2580944}
\BIBentrySTDinterwordspacing

\bibitem{das2007google}
A.~S. Das, M.~Datar, A.~Garg, and S.~Rajaram, ``Google news personalization:
  scalable online collaborative filtering,'' in \emph{Proceedings of the 16th
  international conference on World Wide Web}.\hskip 1em plus 0.5em minus
  0.4em\relax ACM, 2007, pp. 271--280.

\bibitem{middleton2004ontological}
S.~E. Middleton, N.~R. Shadbolt, and D.~C. De~Roure, ``Ontological user
  profiling in recommender systems,'' \emph{ACM Transactions on Information
  Systems (TOIS)}, vol.~22, no.~1, pp. 54--88, 2004.

\bibitem{schwartz2013personality}
H.~A. Schwartz, J.~C. Eichstaedt, M.~L. Kern, L.~Dziurzynski, S.~M. Ramones,
  M.~Agrawal, A.~Shah, M.~Kosinski, D.~Stillwell, M.~E. Seligman \emph{et~al.},
  ``Personality, gender, and age in the language of social media: The
  open-vocabulary approach,'' \emph{PloS one}, vol.~8, no.~9, p. e73791, 2013.

\bibitem{correa2010interacts}
T.~Correa, A.~W. Hinsley, and H.~G. De~Zuniga, ``Who interacts on the web?: The
  intersection of users’ personality and social media use,'' \emph{Computers
  in Human Behavior}, vol.~26, no.~2, pp. 247--253, 2010.

\bibitem{bamman2014gender}
D.~Bamman, J.~Eisenstein, and T.~Schnoebelen, ``Gender identity and lexical
  variation in social media,'' \emph{Journal of Sociolinguistics}, vol.~18,
  no.~2, pp. 135--160, 2014.

\bibitem{you2015picture}
Q.~You, S.~Bhatia, and J.~Luo, ``A picture tells a thousand words--about you!
  user interest profiling from user generated visual content,'' \emph{arXiv
  preprint arXiv:1504.04558}, 2015.

\bibitem{ottoni2014pins}
R.~Ottoni, D.~Las~Casas, J.~P. Pesce, W.~Meira~Jr, C.~Wilson, A.~Mislove, and
  V.~Almeida, ``Of pins and tweets: Investigating how users behave across
  image-and text-based social networks,'' \emph{AAAI ICWSM}, 2014.

\bibitem{lovato2013we}
P.~Lovato, A.~Perina, D.~S. Cheng, C.~Segalin, N.~Sebe, and M.~Cristani, ``We
  like it! mapping image preferences on the counting grid,'' in \emph{Image
  Processing (ICIP), 2013 20th IEEE International Conference on}.\hskip 1em
  plus 0.5em minus 0.4em\relax IEEE, 2013, pp. 2892--2896.

\bibitem{lovato2014faved}
P.~Lovato, M.~Bicego, C.~Segalin, A.~Perina, N.~Sebe, and M.~Cristani, ``Faved!
  biometrics: Tell me which image you like and i'll tell you who you are,''
  \emph{Information Forensics and Security, IEEE Transactions on}, vol.~9,
  no.~3, pp. 364--374, 2014.

\bibitem{gibson1950perception}
J.~J. Gibson, ``The perception of the visual world.'' 1950.

\bibitem{chopra2005learning}
S.~Chopra, R.~Hadsell, and Y.~LeCun, ``Learning a similarity metric
  discriminatively, with application to face verification,'' in \emph{Computer
  Vision and Pattern Recognition, 2005. CVPR 2005. IEEE Computer Society
  Conference on}, vol.~1.\hskip 1em plus 0.5em minus 0.4em\relax IEEE, 2005,
  pp. 539--546.

\bibitem{schifanella15image}
R.~Schifanella, M.~Redi, and L.~M. Aiello, ``An image is worth more than a
  thousand favorites: Surfacing the hidden beauty of flickr pictures,'' in
  \emph{ICWSM'15: Proceedings of the 9th AAAI International Conference on
  Weblogs and Social Media}.\hskip 1em plus 0.5em minus 0.4em\relax AAAI.

\bibitem{yeh2010personalized}
C.-H. Yeh, Y.-C. Ho, B.~A. Barsky, and M.~Ouhyoung, ``Personalized photograph
  ranking and selection system,'' in \emph{Proceedings of the international
  conference on Multimedia}.\hskip 1em plus 0.5em minus 0.4em\relax ACM, 2010,
  pp. 211--220.

\bibitem{zhong2013sharing}
C.~Zhong, S.~Shah, K.~Sundaravadivelan, and N.~Sastry, ``Sharing the loves:
  Understanding the how and why of online content curation.'' in \emph{ICWSM},
  2013.

\bibitem{bernardini2014pin}
C.~Bernardini, T.~Silverston, and O.~Festor, ``A pin is worth a thousand words:
  Characterization of publications in pinterest,'' in \emph{Wireless
  Communications and Mobile Computing Conference (IWCMC), 2014
  International}.\hskip 1em plus 0.5em minus 0.4em\relax IEEE, 2014, pp.
  322--327.

\bibitem{kim2012web}
G.~Kim, L.~Fei-Fei, and E.~P. Xing, ``Web image prediction using multivariate
  point processes,'' in \emph{Proceedings of the 18th ACM SIGKDD international
  conference on Knowledge discovery and data mining}.\hskip 1em plus 0.5em
  minus 0.4em\relax ACM, 2012, pp. 1068--1076.

\bibitem{deng2011hierarchical}
J.~Deng, A.~C. Berg, and L.~Fei-Fei, ``Hierarchical semantic indexing for large
  scale image retrieval,'' in \emph{Computer Vision and Pattern Recognition
  (CVPR), 2011 IEEE Conference on}.\hskip 1em plus 0.5em minus 0.4em\relax
  IEEE, 2011, pp. 785--792.

\bibitem{fu2015tagging}
J.~Fu, T.~Mei, K.~Yang, H.~Lu, and Y.~Rui, ``Tagging personal photos with
  transfer deep learning,'' in \emph{Proceedings of the 24th International
  Conference on World Wide Web}.\hskip 1em plus 0.5em minus 0.4em\relax
  International World Wide Web Conferences Steering Committee, 2015, pp.
  344--354.

\bibitem{danescu2013no}
C.~Danescu-Niculescu-Mizil, R.~West, D.~Jurafsky, J.~Leskovec, and C.~Potts,
  ``No country for old members: User lifecycle and linguistic change in online
  communities,'' in \emph{Proceedings of the 22nd international conference on
  World Wide Web}.\hskip 1em plus 0.5em minus 0.4em\relax International World
  Wide Web Conferences Steering Committee, 2013, pp. 307--318.

\bibitem{taigman2014deepface}
Y.~Taigman, M.~Yang, M.~Ranzato, and L.~Wolf, ``Deepface: Closing the gap to
  human-level performance in face verification,'' in \emph{Computer Vision and
  Pattern Recognition (CVPR), 2014 IEEE Conference on}.\hskip 1em plus 0.5em
  minus 0.4em\relax IEEE, 2014, pp. 1701--1708.

\bibitem{lin2015learning}
T.-Y. Lin, Y.~Cui, S.~Belongie, and J.~Hays, ``Learning deep representations
  for ground-to-aerial geolocalization,'' in \emph{Proceedings of the IEEE
  Conference on Computer Vision and Pattern Recognition}, 2015, pp. 5007--5015.

\bibitem{yang2015CIKM}
L.~Yang, Y.~Cui, F.~Zhang, J.~P. Pollak, S.~Belongie, and D.~Estrin,
  ``Plateclick: Bootstrapping food preferences through an adaptive visual
  interface,'' in \emph{Proceedings of the 24th ACM International Conference on
  Information and Knowledge Management}.\hskip 1em plus 0.5em minus 0.4em\relax
  ACM, 2015.

\bibitem{sun2013hybrid}
Y.~Sun, X.~Wang, and X.~Tang, ``Hybrid deep learning for face verification,''
  in \emph{Computer Vision (ICCV), 2013 IEEE International Conference
  on}.\hskip 1em plus 0.5em minus 0.4em\relax IEEE, 2013, pp. 1489--1496.

\bibitem{zhou2014learning}
B.~Zhou, A.~Lapedriza, J.~Xiao, A.~Torralba, and A.~Oliva, ``Learning deep
  features for scene recognition using places database,'' in \emph{Advances in
  Neural Information Processing Systems}, 2014, pp. 487--495.

\bibitem{krizhevsky2012imagenet}
A.~Krizhevsky, I.~Sutskever, and G.~E. Hinton, ``Imagenet classification with
  deep convolutional neural networks,'' in \emph{Advances in neural information
  processing systems}, 2012, pp. 1097--1105.

\bibitem{ioffe2015batch}
S.~Ioffe and C.~Szegedy, ``Batch normalization: Accelerating deep network
  training by reducing internal covariate shift,'' \emph{arXiv preprint
  arXiv:1502.03167}, 2015.

\bibitem{hadsell2006dimensionality}
R.~Hadsell, S.~Chopra, and Y.~LeCun, ``Dimensionality reduction by learning an
  invariant mapping,'' in \emph{Computer vision and pattern recognition, 2006
  IEEE computer society conference on}, vol.~2.\hskip 1em plus 0.5em minus
  0.4em\relax IEEE, 2006, pp. 1735--1742.

\bibitem{jia2014caffe}
Y.~Jia, E.~Shelhamer, J.~Donahue, S.~Karayev, J.~Long, R.~Girshick,
  S.~Guadarrama, and T.~Darrell, ``Caffe: Convolutional architecture for fast
  feature embedding,'' in \emph{Proceedings of the ACM International Conference
  on Multimedia}.\hskip 1em plus 0.5em minus 0.4em\relax ACM, 2014, pp.
  675--678.

\bibitem{macqueen1967some}
J.~MacQueen \emph{et~al.}, ``Some methods for classification and analysis of
  multivariate observations,'' in \emph{Proceedings of the fifth Berkeley
  symposium on mathematical statistics and probability}, vol.~1, no.~14.\hskip
  1em plus 0.5em minus 0.4em\relax Oakland, CA, USA., 1967, pp. 281--297.

\bibitem{monroe2008fightin}
B.~L. Monroe, M.~P. Colaresi, and K.~M. Quinn, ``Fightin'words: Lexical feature
  selection and evaluation for identifying the content of political conflict,''
  \emph{Political Analysis}, vol.~16, no.~4, pp. 372--403, 2008.

\bibitem{lowe2004distinctive}
D.~G. Lowe, ``Distinctive image features from scale-invariant keypoints,''
  \emph{International journal of computer vision}, vol.~60, no.~2, pp. 91--110,
  2004.

\bibitem{van2008visualizing}
L.~Van~der Maaten and G.~Hinton, ``Visualizing data using t-sne,''
  \emph{Journal of Machine Learning Research}, vol.~9, no. 2579-2605, p.~85,
  2008.

\bibitem{le2014distributed}
Q.~V. Le and T.~Mikolov, ``Distributed representations of sentences and
  documents,'' \emph{arXiv preprint arXiv:1405.4053}, 2014.

\bibitem{park2006naive}
S.-T. Park, D.~Pennock, O.~Madani, N.~Good, and D.~DeCoste, ``Na{\"\i}ve
  filterbots for robust cold-start recommendations,'' in \emph{Proceedings of
  the 12th ACM SIGKDD international conference on Knowledge discovery and data
  mining}.\hskip 1em plus 0.5em minus 0.4em\relax ACM, 2006, pp. 699--705.

\end{thebibliography}

\end{document}